# Generate labeled training data using Prompt Programming and GPT-3. An example of Big Five Personality Classification

Eason Chen
Learning Sciences
National Taiwan Normal University
eason.tw.chen@gmail.com

## Abstract

We generated 25000 conversations labeled with Big Five Personality traits using prompt programming at GPT-3. Then we train Big Five classification models with these data and evaluate them with 2500 data from generated dialogues and real conversational datasets labeled in Big Five by human annotators. The results indicated that this approach is promising for creating effective training data. We then compare the performance by different training approaches and models. Our results suggest that using Adapter-Transformers and transfer learning from pre-trained RoBERTa sentiment analysis model will perform best with the generated data. Our best model obtained an accuracy of 0.71 in generated data and 0.65 in real datasets. Finally, we discuss this approach's potential limitations and confidence metric.

## 1   Introduction

The traditional process of annotating training data has been time-consuming and requires significant human resources. Nevertheless, the advent of large Natural Language Generation (NLG) models such as GPT-3 (Brown et al., 2020) has made it possible to generate controllable outputs through prompts. In this paper, our research question is 'can we generate effective labeled training data by GPT-3?'

To answer the question, we generate dialog messages with a specific Big Five personality trait label by prompt programming. Then, based on these data, we evaluate the model training results with human annotators in both generated and real conversational data. Moreover, we compared different classification models training approaches and pre-trained models with the generated labeled data. The goal is to achieve prediction accuracy that are closest to human-labeled. Lastly, we discussed potential applications and limitations of using generated data to predict real-world datasets.

We hope our result can boost natural language processing research on topics lacking existing dataset and even languages need more linguistic data. Future researchers don't need to annotate data manually anymore, but can let NLG models do the job. By doing so, researchers can focus more on collecting and analyzing data to get insights.

## 2   Relative Work

### 2.1   Prompt Programming

Natural language generation (NLG) uses machine learning to produce natural language (Radford et al., 2019). NLG is done by analyzing the given input and using it to generate human language resembled output that accurately reflects the input (Holtzman, Buys, Du, Forbes, & Choi, 2019). For large language model which trained with lots of data, it is possible to control the output by tunning the given input or add additional prompts before the input. This method is so call Prompt programming.

Prompt programming (Reynolds & McDonell, 2021) is a technique that utilizes specific input, or prompts, to guide the output generated by NLG models, allowing for greater control over the produced data. For example, in Table 1 for the dialog generation scene, given the same message input: 'The boss keeps making things difficult for me.' The generated reply (in bold text) from friend, who is powered by GPT-3, will vary according to the prompt (underline) provided before the input.

The friend in example 1 is more open-minded while the friend in example 2 is less open-minded. Changing the prompt can cause the generated responses of the friend to diverge.

In this paper, we tried to generate labeled training data for dialogue messages through different prompts, which can then be used to predict the attributes of the message sender. The attribute we decided is Big Five Personality traits.

| (Example 1) | |
|---|---|
| 1 | The following is your conversation with your friend, who is <u>intellectual, imaginative, sensitive, and open-minded.</u> |
| 2 | You: The boss keeps making things difficult for me. |
| 3 | Friend: **What can you do to change the situation?** |
| (Example 2) | |
| 1 | The following is your conversation with your friend, who is <u>down-to-earth, insensitive, and conventional.</u> |
| 2 | You: The boss keeps making things difficult for me. |
| 3 | Friend: **That's just the way the world works. You can't always get what you want.** |

Table 1: Example of NLG with prompts for dialog message generation.

## 2.2 Big Five Personality Traits

Big Five Personality (Roccas et al., 2002) is a set of factors that describe most personality traits: Neuroticism (NEO), Extroversion (EXT), Openness To Experience (OPE), Agreeableness (AGR), and Conscientiousness (CON). Big Five is useful in inference people's other attributes, such as work performance (Zhao & Seibert, 2006), social strategy (Agbaria & Mokh, 2022), personal value (Roccas et al., 2002) or even political stance (Gerber, Huber, Doherty, & Dowling, 2011).

Researcher so far are dedicated in predicting peoples' Big Five Personality (Skowron, Tkalčič, Ferwerda, & Schedl, 2016; Tandera, Suhartono, Wongso, & Prasetio, 2017). Nevertheless, no public natural language datasets labeled in the Big Five have been found yet. Hence, generating data labeled in Big Five will be quite meaningful.

## 3 Methods

### 3.1 Generating dialogue in Big Five

We recruited 120 participants in Taiwan from the internet through convenience sampling with a mean age of 24.7 years (range 18 – 64, SD = 7.9). Participants are asked to chat 10 turns with 20 dialog agents respectively who designed with different Big Five characters.

Total 20 chat agents are designed in 5 (Big Five, ex: Extraversion) x 2 (opposite of Big Five, ex: Introversion) x 2 (Gender). The Big Five Personality prompts are created based on the adjective description from Roccas et al. (2002). Examples of generating agent's replies in different character by prompts can be found in Table 1. Given the description, the prompt before the input will be "The following is your conversation with your friend, who is [description]." Example can be found at Table 1 at section 2.1. All description are Big Five definition from Roccas et al. (2002).

| Personality | Description |
|---|---|
| Neuroticism | Anxious, depressed, angry, and insecure |
| Non-Neuroticism | Calm, poised, and emotionally stable. |
| Openness | Intellectual, imaginative, sensitive, and open-minded. |
| Non-Openness | down-to-earth, insensitive, and conventional. |
| Agreeableness | good-natured, compliant, modest, gentle, and cooperative. |
| Non-Agreeableness | irritable, ruthless, suspicious, and inflexible. |
| Conscientiousness | careful, thorough, responsible, organized, and scrupulous. |
| Non-Conscientiousness | irresponsible, disorganized, and unscrupulous. |
| Extroversion | sociable, talkative, assertive, and active. |
| Non-Extroversion | retiring, reserved, and cautious. |

Table 6: Big Five Personality description we used from Roccas et al. (2002)

### 3.2 Training the model

We use *bert-base-cased* (Devlin, Chang, Lee, & Toutanova, 2018) as the baseline model to compare different training approaches. Then after decide the best practice, we compare the performance of different pre-trained models, including RoBERTa (Liu et al., 2019) and sentiment analysis model.

We investigated three approaches: together, separate, and Adapter-Transformers. Together (TO) is to train a classification model with all labels. For example, when training Big Five traits, as shown in figure 1, the output dimension will be 10, which is 5 (Big Five) + 5 (opposite). The advantage of this approach is simple as we only need to train one model. Yet the downside of this approach is that when training, each label will tend to be gradient descent to as low as possible, leading to overfitting.

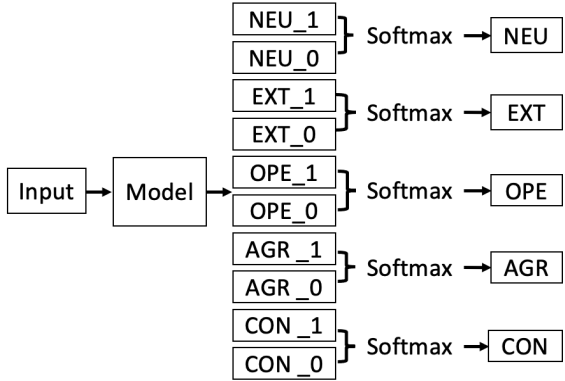

Figure 1: Training all labels together as classification model.

The second approach is separately (SE). That is, we trained labels respectively on their own model with binary classification. For example, the Openness label will have its' own model with two outputs, Openness and Non-Openness (Figure 2). The advantage of this approach is less possible to overfit by unrelated training data. Nevertheless, it is more complicated to train and manage five models than one. For example, with Big Five traits, the storage size, memory consumption, and prediction time will be multiplied five times.

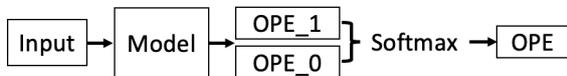

Figure 2: Training one model per label

Lastly, we applied Adapter-Transformers (AT) (Pfeiffer et al., 2020). Adapter Transformers is a technique for fine-tuning pre-trained transformer models. It involves adding trainable parameters, so called adapters, to the frozen pre-trained model, then fine-tuned those additional parameters with the task-specific data. Adapter-Transformers includes the advantages and eliminates the disadvantages of the former two. We only need to store five tiny adapters and can train each label separately. Moreover, we can use the parallel output feature to predict all labels together (Fig. 3).

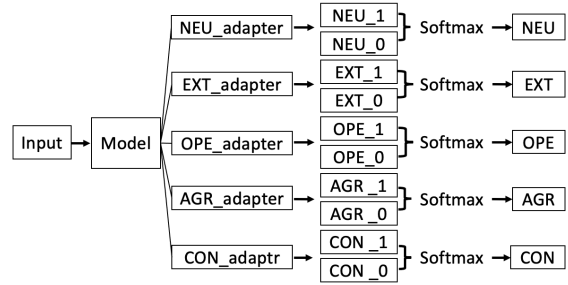

Figure 3: Training Big Five Traits with Adapter-Transformers

### 3.3 Training and Testing Dataset

We randomly separated 1000 outputs from the generated 25000 training data. Moreover, we randomly choose additional 1500 data from three real world dialog dataset, which include real conversation in various domain and topics. The Datasets we choose are MultiWOZ (Zang et al., 2020), ConvAI (Logacheva, Burtsev, Malykh, Polulyakh, & Seliverstov, 2018), and Cornell Movie-Dialogs Corpus (Danescu-Niculescu-Mizil & Lee, 2011).

### 3.4 Human annotation

We recruit 4 data annotators from Taiwan through convenience sampling with a mean age of 26 (range 21 – 32) who're professional in English and familiar with Big Five. For each testing data, they will assess the message sender's level of Big Five traits and difficulty in making decisions on a scale from 1 to 10.

## 4 Results

In this section, we evaluate the model performance trained by generated dialogue. We use pandas to preprocess the data. All models are trained in PyTorch for 50 epochs with a batch size of 32 on RTX 3090 on servers with two Intel(R) Xeon(R) Gold 6230 CPU @ 2.10GHz and 128GB RAM. Source codes, data, and predictions are provided in the supplementary material.

### 4.1 Evaluation on Generated data

The accuracy of different training approach on the generated data, from high to low, is using Adapter-Transformers (AT, acc = 0.701), training separately (SP, acc = 0.62) and together (TO, acc = 0.6).

We then further explore the performance of different pre-trained models. The results indicated that the accuracy of RoBERTa (acc = 0.704) is slightly better than BERT. Moreover, if we applied

RoBERTa model pre-trained by twitter sentiment analysis from Loureiro, Barbieri, Neves, Anke, and Camacho-Collados (2022), the performance will be even better (acc = 0.71).

| Architectures | Accuracy on Big Five Traits with Generated Data | | | | | Avg |
|---|---|---|---|---|---|---|
| | EXT | AGE | OPE | CON | NEU | |
| BERT-base (TO) | 0.586 | 0.617 | 0.611 | 0.551 | 0.629 | 0.598 |
| BERT-base (SE) | 0.618 | 0.724 | 0.654 | 0.534 | 0.605 | 0.627 |
| bert-base (AT) | 0.689 | 0.799 | 0.733 | 0.571 | 0.714 | 0.701 |
| RoBERTa-base (AT) | 0.644 | 0.831 | 0.761 | 0.586 | 0.701 | 0.704 |
| RoBERTa-sentiment (AT) | 0.688 | 0.822 | 0.77 | 0.547 | 0.725 | 0.710 |

Table 2: Evaluation result with generated data

## 4.2 Evaluation on real-world data

We evaluate the model with real-world dataset. The results show each model has its own merits. For the pre-trained sentiment RoBERTa with AT, model performances from high to low are Movie-Dialogs Corpus (acc = 0.671), MultiWOZ (acc = 0.658), and ConvAI (acc = 0.620). Nevertheless, BERT with AT outperform others at multiOz (acc = 0.686) and convAI (acc = 0.621).

| Model | Movie | multiOz | convAI | Avg |
|---|---|---|---|---|
| BERT-base (TO) | 0.585 | 0.598 | 0.529 | 0.571 |
| BERT-base (SE) | 0.606 | 0.575 | 0.560 | 0.581 |
| BERT-base (AT) | 0.664 | 0.686 | 0.621 | 0.656 |
| RoBERTa-base (AT) | 0.630 | 0.615 | 0.600 | 0.615 |
| RoBERTa-sentiment (AT) | 0.671 | 0.658 | 0.620 | 0.650 |

Table 3: Evaluation result with real dataset

## 4.3 Performance on each Big Five dimension

We examined the performance on each Big Five personality traits. The results indicated that, for the pre-trained sentiment RoBERTa with AT, model performances from high to low are Agreeableness (acc = 0.75), Neuroticism (acc = 0.701), Extroversion (acc = 0.678), Openness to Experience (acc = 0.664), and Conscientiousness (acc = 0.532). Nevertheless, the accuracy of BERT with AT on Conscientiousness (acc = 0.58) outperform others.

| Architectures | Accuracy on Big Five Traits with Generated Data | | | | | Avg |
|---|---|---|---|---|---|---|
| | EXT | AGE | OPE | CON | NEU | |
| BERT-base (TO) | 0.554 | 0.6 | 0.56 | 0.549 | 0.626 | 0.578 |
| BERT-base (SE) | 0.574 | 0.642 | 0.599 | 0.542 | 0.604 | 0.592 |
| BERT-base (AT) | 0.666 | 0.756 | 0.639 | 0.58 | 0.698 | 0.668 |
| RoBERTa-base (AT) | 0.578 | 0.730 | 0.635 | 0.577 | 0.667 | 0.638 |
| RoBERTa-sentiment (AT) | 0.678 | 0.75 | 0.664 | 0.532 | 0.701 | 0.665 |

Table 4: General accuracy on each Big Five dimension

## 4.4 Relation between Difficulty and Output

A Pearson correlation coefficient was computed to assess the linear relationship between the annotators' perceived difficulty and the processed output, which is the absolute value of raw outputs' positive value minus negative value. For example, if a model output in Openness's negative and positive is -0.1 and -0.8, then the processed output will be 0.9. Yet if it is 0.1 and 0.1, the processed output will be 0.2. The results indicated that, in general, the processed output is negatively correlated to the difficulty.

| Architectures | linear relationship between the annotators' perceived difficulty and the processed output | | | | |
|---|---|---|---|---|---|
| | EXT | AGE | OPE | CON | NEU |
| BERT-base (AT) | -.12*** | -.19*** | -.21*** | -.17*** | -.26*** |
| RoBERTa-base (AT) | .005** | -.25*** | -.35*** | -.13*** | -.34*** |
| RoBERTa-sentiment (AT) | .03 | -.12*** | -.17*** | -.15*** | -.17*** |

Note: *** p < .001, ** p < 0.01

Table 5: Relation between difficulty and output

## 5 Discussion and Limitation

Our best model obtained an accuracy of 0.71 in generated data and 0.65 in the real datasets. Although the best prediction about the Big Five from previous research is 0.74 (Tandera et al., 2017), their model references many personal data and posts on users' Facebook. Therefore, as our models only predicted by a single message and trained by generated data, the results are promising.

Our model's accuracy in real-data is worse than generated output. This might be because real-data includes ambiguous dialogue. For example, it is hard to tell the message sender's Big Five by "I love meat. I eat lots of meat!" Moreover, real-data includes topic related to news and trends that can be recognized by human labeler. However, our data is generated from GPT-3, which data is only until June 2021. That is, we may not be able to generate up-to-date content and might influence the performance in real-data. Therefore, it is vital to keep in mind that a single message is not enough to make a sound prediction. It will be interesting to apply this model with many messages from the same message sender and then average the output.

Our results also showed that the processed output is negatively correlated to the labelers' perceived difficulty on judgement. Therefore, this metric might can be used as the "confidence level" of the given input. For example, if the predicted output for Neuroticism from "I'm so depressing!" is [-0.3, 0.9], the processed output will be 1.2. If the output from "I love meat" is [0.1, 0.2], the processed output will be 0.3. Even though they both belong to Neuroticism, their confidence level will be different.

Lastly, the amount and quality of prompts to generate labeled training data are essential. The performance might be even better if we use many different prompts for each Big Five dimension.

# 6 Conclusion

We research generating Big Five labeled training data using prompt programming at GPT-3. The results indicated that this approach is promising for creating effective training data. Moreover, our results suggest that using Adapter-Transformers and transfer learning from pre-trained RoBERTa sentiment analysis model will perform best. Our best model trained by generated data obtained an accuracy of 0.71 in generated data and 0.65 in real datasets.

## Ethics Statement

The study was approved by the Institutional Review Board of (anonymous) with protocol code (anonymous).

## Acknowledgments

This work was supported by (anonymous) under Grants (anonymous) and (anonymous).

status: a meta-analytical review. *Journal of applied psychology, 91*(2), 259.

# A Prompts used to generate messages with Big Five Personality

Given the description, the prompt before the input will be "The following is your conversation with your friend, who is [description]." Example can be found at Table 1 at section 2.1. All description are Big Five definition from Roccas et al. (2002).

| Personality | Description |
|---|---|
| Neuroticism | Anxious, depressed, angry, and insecure |
| Non-Neuroticism | Calm, poised, and emotionally stable. |
| Openness | Intellectual, imaginative, sensitive, and open-minded. |
| Non-Openness | down-to-earth, insensitive, and conventional. |
| Agreeableness | good-natured, compliant, modest, gentle, and cooperative. |
| Non-Agreeableness | irritable, ruthless, suspicious, and inflexible. |
| Conscientiousness | careful, thorough, responsible, organized, and scrupulous. |
| Non-Conscientiousness | irresponsible, disorganized, and unscrupulous. |
| Extroversion | sociable, talkative, assertive, and active. |
| Non-Extroversion | retiring, reserved, and cautious. |

Table 6: Big Five Personality description we used from Roccas, Sagiv, Schwartz, and Knafo (2002)